# Virtual Memory Partitioning for Enhancing Application Performance in Mobile Platforms

Geunsik Lim, *Student Member*, IEEE, Changwoo Min, and Young Ik Eom

**Abstract** — Recently, the amount of running software on smart mobile devices is gradually increasing due to the introduction of application stores. The application store is a type of digital distribution platform for application software, which is provided as a component of an operating system on a smartphone or tablet. Mobile devices have limited memory capacity and, unlike server and desktop systems, due to their mobility they do not have a memory slot that can expand the memory capacity. Low memory killer (LMK) and out-of-memory killer (OOMK) are widely used memory management solutions in mobile systems. They forcibly terminate applications when the available physical memory becomes insufficient. In addition, before the forced termination, the memory shortage incurs thrashing and fragmentation, thus slowing down application performance. Although the existing page reclamation mechanism is designed to secure available memory, it could seriously degrade user responsiveness due to the thrashing. Memory management is therefore still important especially in mobile devices with small memory capacity.

This paper presents a new memory partitioning technique that resolves the deterioration of the existing application life cycle induced by LMK and OOMK. It provides a completely isolated virtual memory node at the operating system level. Evaluation results demonstrate that the proposed method improves application execution time under memory shortage, compared with methods in previous studies[1].

## I. INTRODUCTION

Recent mobile devices support both built-in and downloaded applications from application stores [1], [2]. The application store is a type of digital distribution platform designed to release application software. Memory management in mobile devices is still very important because the devices have relatively small memory capacity with no ad-hoc expansion, and the memory management of downloaded applications cannot be controlled or tested at the time of manufacturing. Therefore, memory shortage is likely to occur more frequently. To cope with the memory shortage, low memory killer (LMK) [1], [3]-[5] is the most widely adopted solution. Under memory shortage, it repeatedly terminates less important applications in a forceful way until the operating system (OS) secures enough free memory space to run a new application. The list of the order of application importance is managed by user-space daemons, such as a thread manager and an activity manager. The activity manager acts as a traffic controller for the overall activities (e.g. foreground and background activities and system resources) running on the mobile device. The goal of the activity manager is to balance activity priorities and system resources to optimize the user's experience. The frequent operations of LMK and out-of-memory killer (OOMK) could seriously deteriorate user-perceived performance in two ways. First, because all relevant memory space of a victim application [6] is unloaded, the unloaded memory should be reloaded at the next launching of the victim application, and it could seriously slow down the application performance. To select a victim application, OS considers the following criteria: the number of threads, the central processing unit (CPU) running time, the scheduling priority, and whether or not it directly accesses the hardware. Second, the core built-in applications, such as Phone, short message service (SMS), and Contacts, can be forcibly terminated.

When page faults induced by the memory shortage occur frequently, the cost of page replacement dominates CPU utilization, making applications more prone to miss the required deadline [7], [8]. As a result, instead of actually obtaining free memory, the *thrashing* [9] frequently occurs. Consequently, a user encounters slow performance even in built-in applications. In this paper, the proposed techniques support new memory partitioning at the OS level, which limits the page reclamation within the partitioned memory range based on the well-defined hierarchy importance of applications. The hierarchy of applications is classified into built-in applications, applications from trusted sources, and unknown applications from untrusted sources.

The remainder of this paper is organized as follows. Section II describes the memory management problems of the existing mobile platform. Section III addresses the design and implementation of the proposed techniques. Section IV shows the evaluation results. Related work is described in Section V. Finally, Section VI concludes the paper.

[1] This work was supported by the IT R&D program of MKE/KEIT [10041244, Smart TV 2.0 Software Platform]. This research was supported by Basic Science Research Program through the National Research Foundation of Korea (NRF) funded by the Ministry of Education, Science and Technology (2010-0022570).

Geunsik Lim, Changwoo Min, and Young Ik Eom are with the College of Information and Communication Engineering, Sungkyunkwan University (SKKU), 300 Cheoncheon-dong, Jangan-gu, Suwon, Republic of Korea. (e-mail: {leemgs, multics69, yieom}@skku.edu).

## II. MEMORY MANAGEMENT IN MOBILE PLATFORMS

In this section, the most widely used memory management features, including page reclamation, swap in/out, process container, LMK, and OOMK, to secure free memory under memory pressure will be presented.

### A. Conventional Memory Management

The page reclamation mechanism [10]-[12] is useful to obtain the available memory space on the system. However, it finds target pages that are sacrificial in memory reclamation, based on the *least recently used page* (LRU) replacement algorithm [6], [13]. It blindly handles all processes without the platform level semantics, which are important system applications in a mobile platform.

The swap in/out mechanism [14] is widely used to run applications that require larger memory than the physical memory capacity. Unfortunately, most of the mobile device manufacturers do not use the swap in/out mechanism [15]-[17]. Because swapping operations work with a slow storage device with limited endurance, they fail to provide reasonably predictable performance [18]-[20].

The process container [21]-[23], also called the resource controller, manages hierarchically organized process groups. It controls the resource usage of process groups by limiting the sum of memory usage in a group. It cannot resolve memory fragmentation or provide memory isolation because it only logically partitions memory space by using a *per-group least recently used page* list [24]. Therefore, it cannot isolate the address access of the physical memory because the page reclamation executes in a unified flat memory.

### B. Memory Management of LMK

The existing mobile platforms manage the memory management of the applications in a single memory space. These applications mainly consist of built-in applications by the manufacturer and external applications downloaded from the application store by the user. The original role of LMK is to automatically terminate the applications in an LRU list [1], [3]-[5] when the available memory reaches a specified threshold of the system. The operating system starts to kill the oldest unneeded processes in the LRU list to retrieve the free memory space for the execution of new applications. If the system reaches the threshold of free physical memory, LMK terminates the applications that are relatively less important among the running applications.

However, the memory fragmentation gradually increases because the operating system reclaims the memory blocks of unimportant processes with the unit of page from a physical memory. As the memory fragmentation becomes more severe, the small size memory blocks increase further, resulting in additional memory management costs such as the merging of small blocks by a memory allocator [25], the time required to read all the nonadjacent memory blocks at once, and the scheduling cost between the memory blocks because there are too many small blocks. The many small blocks increase the memory scheduling cost to determine whether the higher priority processes are waiting or running during the allocation and the release of the small blocks in the *preemptive operating system*. For example, in the case of systems such as a camcorder, which requires large IO operations, releasing many small memory blocks is time consuming [26].

In the mobile environment, user responsiveness is more important than the fairness of the task, contrary to the server environment. Previous studies concentrate on killing applications in the same memory area when the available memory is insufficient. The proposed idea focuses on how to execute the page reclamation in an isolated memory area to solve the performance slowdown of time-critical trusted applications when the physical memory reaches memory shortage. It prevents the performance slowdown of the trusted applications from *thrashing* [9] that could occur by the indiscreet memory usage of untrusted applications [27].

### C. Memory Management of OOMK

OOMK [28], [29] endeavors to overcome the memory shortage from the out-of-memory status by terminating a lower priority process. The original role of OOMK is to kill unimportant processes based on the memory score of processes heuristically when the memory capacity is deficient. However, the operation of OOMK seriously degrades the execution speed of new applications due to the *thrashing* [9]. When a new application is launched under a high memory pressure, OOMK forcibly terminates a process based on the relative severity in order to retrieve additional memory space. OOMK attempts to retrieve the available memory by killing the processes of the lower memory score [12] as a victim process to avoid an out-of-memory situation. It heuristically determines the victim processes according to the number of execution frequencies of the application, the execution time of the application, the scheduling priority of the process, the application to access devices, and the application authorized by the root user.

Although additional physical memory space settles the memory shortage situation, the depletion status of memory [30] frequently occurs whenever users try to run massive applications or memory-intensive applications. Moreover, the miniaturization of the device, the reduction of production cost, and the minimization of power consumption are very important to mobile device manufacturers. Therefore, system software technology to manage the memory consumption of user-space applications is significant.

In summary, the existing mobile platforms provide various kernel features for running new applications under the memory shortage: conventional memory management schemes, memory management of LMK, and memory management of OOMK. However, these kernel features frequently induce thrashing, page fault, and page replacement, and thus result in low speed applications. It is very important that the mobile devices are used to run the user application instantly. Especially, core built-in applications must be executed within the specified responsive time, even though the available size of the physical memory is very short.

## III. VIRTUAL MEMORY NODES TO AVOID LMK/OOMK OPERATIONS

This section describes the design and implementation of the virtual memory partitioning framework at the operating system level to solve the problem of the lack of available physical memory space of the application that results in poor user responsiveness of the time-critical core applications. This technique sets the memory layout dynamically which is based on the permission privilege of the root user at boot time to support the various mobile devices from the low-end to the high-end. This technique is named the Virtual Memory Node (VNODE) [3].

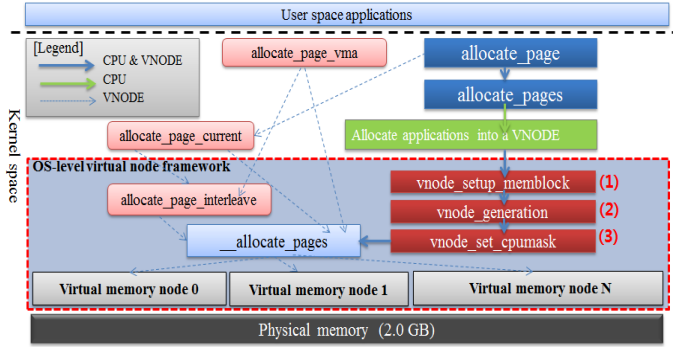

**Fig. 1. Architecture of the virtual memory partitioning scheme**

### A. Design of VNODE

Figure 1 shows the overall architecture of a new memory partitioning technique for the process life cycle of mobile platform applications that have limited physical memory space. The process life cycle of the mobile platform controls the status of the processes for the user responsiveness of the application. The proposed memory partitioning technique mainly consists of three components as follows:
1) *vnode_setup_memblock,* which manages the mapping between the physical memory address and a virtual node to separate the physical memory.
2) *vnode_generation*, which generates the specified virtual memory nodes from a physical memory node and determines the size of the table for holding the address range of the physical memory.
3) *vnode_set_cpumask*, which allocates the CPU masks to support mapping between a virtual memory node(s) and specified CPU(s) to recognize CPU-Hotplug and CPU-DVFS enabled multicore environments [31].

VNODE has two main advantages for mobile devices with limited memory capacity as follows:
1) Complete memory isolation: VNODE controls unnecessary memory consumption of untrusted applications [27], [32] by splitting a physical memory. For example, VNODE0 for trusted applications such as a built-in package and VNODE1 for untrusted applications such as mal-ware software and memory hog software.
2) Reduces the number of LMK/OOMK operations: VNODE minimizes the possibility of LMK/OOMK operations whenever memory shortage occurs.

The key idea is to allocate and release the memory area of the application in the physically specified memory area to control unknown applications from the untrusted sources. These operations help to avoid the problem of reaching the absence of available memory as soon as possible. Memory features such as on-demand paging, page reclamation, and page defragmentation execute memory allocation and release of the applications [32]. This means that the core built-in applications meet the factors of performance degradation more often due to the unknown applications from the untrusted sources, as follows:
1) Thrashing: harms the execution time of applications because of the page fault and page replacement [7].
2) Memory fragmentation: increases the cost of maintaining too many small memory blocks. It increases the scheduling cost while allocating/releasing the scattered small memory blocks.

The operating system can reduce the cost of the *trashing* and the *fragmentation* of the physical memory via the proposed idea. As a result, the virtual memory partitioning scheme protects untrusted applications from harming the execution time of the time-critical core applications. The proposed system is the complete concept of the virtual memory nodes at the operating system level for this purpose. This feature provides a scalable infrastructure to support mobile devices for various purposes: memory space that is virtually separated from a physical memory, memory isolation at the operating system level, enhanced page reclamation based on virtual memory node, and dynamic memory-controlling interface based on discretionary access control to set up at boot time.

### B. Implementation Details

The existing operating system adopts a flat memory model that allocates and reclaims pages from a single unified memory region to handle the memory resource. The operating system cannot settle memory fragmentation and page reclamation completely because the existing memory subsystem allocates/releases the memory space of a process using a global LRU list. The proposed system supports the virtual memory nodes that are divided into two or more spaces from a physical memory. The virtual memory node isolates the memory usage of the process selectively by controlling the *page table* for each process for the virtual memory scheme.

However, the existing system cannot determine a page boundary region to reclaim pages because the existing approach maintains the memory's usage based on the amount of memory of the processes without the virtual memory access area. Therefore, the operating system can be equipped with a mechanism to allocate the memory pages of the processes such as the virtual memory nodes that appear to be a physical memory.

The *allocate_page_vma* function shown in Figure 1 manages the pages of the applications in the virtual memory space. It connects the memory pages of the process to the *allocate_page_interleave* function.

The *allocate_page_interleave* function executes the low-level operation to interconnect an application and a memory area. If the operating system needs to find the allocated memory address currently according to the process request, the *allocate_page* function calls the *allocate_page_interleave* function via the *allocate_page_current* function.

Finally, the *__allocate_pages* function allocates/releases the memory area of the process using the processing result of VNODE's three components: 1) *vnode_setup_memblock*, 2) *vnode_generation*, and 3) *vnode_set_cpumask*. Table I describes the meaning of the acronyms in the legend of Figure 1.

TABLE I
THE MEANING OF THE ACRONYMS IN THE LEGEND OF FIGURE 1

| LINE NAME | DESCRIPTION |
| --- | --- |
| CPU | This line is an interface for multi-core environments. When the status of CPU is online (or offline), operating system calculates the number of the actual CPUs and the number of the online CPUs. |
| CPU& VNODE | This line indicates a relation view linking the multi-core CPU and virtual memory node(s) for the processes. |
| VNODE | This line expresses the connected point and the relationship among the kernel-level functions when the three key components of VNODE allocate/release memory block after recognizing the status of multi-core CPU. |

*C. Memory Allocation and Page Reclamation for Built-in Applications and External Applications*

The arrows in Figure 1 show the operating structure between CPU and memory. The root user can adjust the generation procedure of the virtual memory nodes at boot time. For example, it will be assumed that for the memory layout, the trusted applications can run in VNODE0 and the untrusted applications can run in VNODE1. In the mobile devices, the definition of the typical two types of software is as follows:
1) Trusted applications: which are the built-in applications and downloaded applications from trusted sources.
2) Untrusted applications: which are downloaded applications from untrusted sources. Untrusted applications [27], [32] potentially include malicious code, memory hog, high power consumption, and unnecessary CPU usage. Abnormal system behavior and system reboot mostly results from these applications.

Through the proposed approach, the operating system controls the applications to avoid reaching memory shortage while running the applications. The proposed memory partitioning technique settles the problem of the single memory space by running the trusted applications within VNODE0 only. That is, the built-in applications from the trusted sources stay in the memory until users directly exit their applications, as shown in Figure 2.

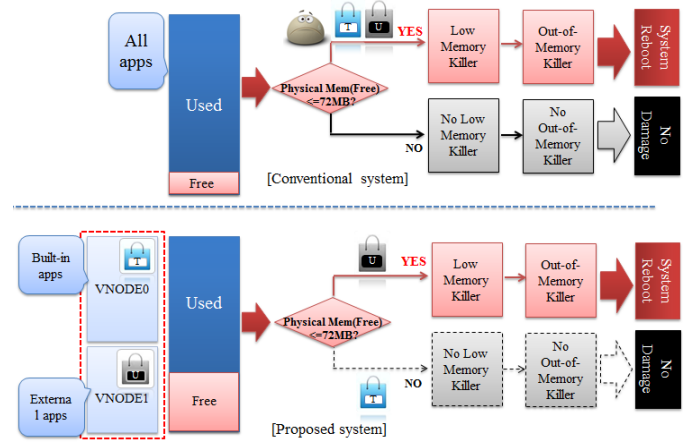

Fig. 2. This flow diagram describes the operation of LMK/OOMK for the trusted applications and the untrusted applications. The external applications are the untrusted applications that run in VNODE1. 'T' refers to the trusted applications of the official application store, and 'U' refers to the untrusted applications of the unofficial application store. The threshold of free physical memory is 72 MB.

*D. Dynamic Memory-Controlling Interface*

Mobile devices are used in many different ways and each device has different system requirements such as the number of built-in applications, the memory capacity, and the clock speed of the CPU. VNODE consists of a dynamic memory-controlling interface to resolve the problem of the lack of shared memory that occurs when dividing a physical memory into two or more virtual nodes.

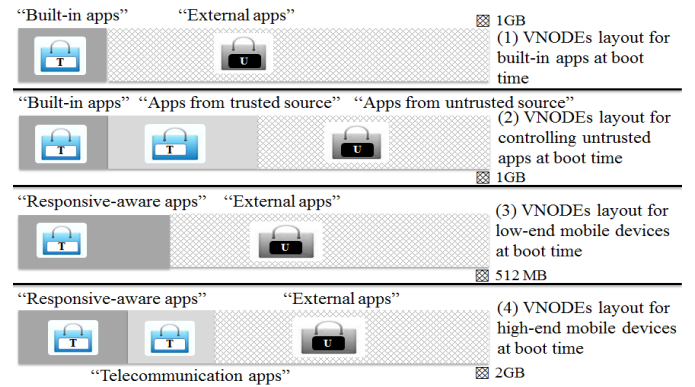

Fig. 3. Cases of the dynamic virtual memory layout at boot time to support various mobile devices. 'Apps' is the abbreviation of applications. '▨' refers to the physically limited memory size.

VNODE includes a dynamic memory-controlling interface via the boot parameter interface to dynamically control the intended virtual memory layout at boot time. This function helps VNODE to work for mobile devices with a variety of characteristics. The dynamic setting of the memory layout at boot time is only permitted according to the permission based security model. Figure 3 shows four examples of smart mobile devices that support application stores:

**Case 1:** VNODE dynamically generates virtual memory nodes at boot time to isolate external applications downloaded from the application store and the built-in applications. It protects the built-in applications from the LMK and OOMK operations to assure the user responsiveness of the built-in applications. The setting of the virtual memory layout is valid until system reboot.

**Case 2:** VNODE generates three virtual memory nodes for built-in applications, the applications from the trusted application store, and the unknown applications from the untrusted application store. This memory layout additionally protects the trusted application store against **Case 1**.

**Case 3:** VNODE generates two virtual memory nodes for responsive-aware applications and external applications for low-end mobile devices including low memory capacity. Low-end devices rapidly reach a memory shortage situation against high-end devices. This memory layout policy assures the execution time of time-critical applications with responsiveness and real-time characteristics, which is always important for Phone, SMS, and Contacts.

**Case 4:** VNODE generates three virtual memory nodes for high-end mobile devices that have a high memory capacity. It consists of three virtual nodes: the responsive-aware built-in applications area provided by the manufacturer, the built-in applications area provided by the telecommunication company, and the large external applications area available due to the sufficient physical memory capacity.

## IV. EVALUATION

This section shows the effect of the approach on the real mobile device including dual-core CPU, 2 GB RAM, 200 GB HDD via SATA interface, and Linux kernel 3.0.8 (Figure 4). Finally, the enhanced existing operating system based on a virtual memory node improves the application performance with the virtual memory partitioning technique.

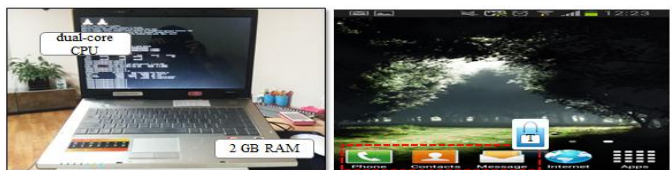

**Fig. 4. Experimental system**

To evaluate the performance efficiency of the proposed techniques, the experimental system was configured to have two virtual nodes at boot time: VNODE0 of 0.5 GB for trusted built-in applications and VNODE1 of 1.5 GB for untrusted external applications. The size of each VNODE could easily be determined by the memory requirement of trusted built-in applications such as Phone, SMS, and Contacts. Since the built-in applications were developed and tested by manufacturers with a specific memory requirement, the experimental system was established based on the memory requirement of built-in applications to the size of VNODE0 and that of VNODE1 for the downloaded applications from the application store. After the booting procedure is completed, test applications are performed to increase the memory pressure on VNODE1. The two test applications were developed to increase the memory pressure in a real mobile device as follows:

1) Increase utilization of page cache: A file with 2 GB is repeatedly read in a sequential order. The page cache is filled up and the memory pressure increases.
2) Increase utilization of anonymous page: This test application repeatedly allocates heap memory with 1.2 GB and fills with zero. It increases the size of anonymous pages and the running process to be selected tends to become a victim process according to page reclamation [7].

Typical usage of the mobile platform was performed while running the two test applications for two days. After running the mobile platform for two days, the system was analyzed according to the following four aspects:

1) Available memory to evaluate an instant execution of application without LMK/OOMK.
2) The number of execution frequency of LMK/OOMK required to verify the assurance of the execution time of the built-in applications.
3) Effect of memory defragmentation to reduce the cost of memory management.
4) The execution time of applications and breakdown of the three experimental effects.

### A. Available Memory

Figure 5 compares the result of the memory consumption between the existing system (before) and the proposed system (after). From the experiments, the free memory is 31 MB in the existing system and 187 MB in the proposed system. The mobile device gained the additional memory of 156 MB compared to the existing system by strictly controlling the unnecessary memory consumption of untrusted applications.

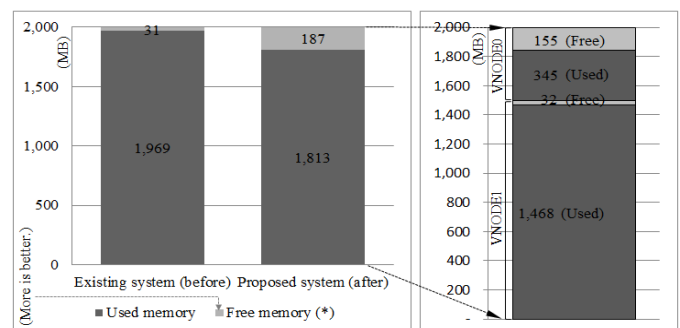

**Fig. 5. Comparison of free memory space between the existing system (before) and the proposed system (after). The graph on the right shows free memory space in two parts corresponding to the two virtual nodes.**

The application allocated in the VNODE1 memory node can only be run in the physical memory of 1.5 GB. Therefore, the operating system attempts to execute the operation of the page reclamation only in the VNODE1 memory node when the available memory is not sufficient. As a result, the proposed system settled the cost of application loading time to transfer to the memory from the storage [33] to run the application by protecting the aggressive LMK and OOMK operations using the novel virtual memory partitioning scheme.

*B. Number of Execution Frequencies of LMK/OOMK*

This section compares the experimental result of the number of execution frequencies of LMK/OOMK between the existing system (before) and proposed system (after). The proposed system dramatically removed the number of execution frequencies of naive memory reclamation [4]-[6] running the LMK operation and OOMK operation sequentially. The experimental result shows that untrusted applications do not increase the number of LMK operations by isolating the page reclamation of the untrusted applications executed only in the VNODE1 of the 1.5 GB memory node. Figure 6 shows the experimental result of the LMK operation: 36 times in the existing system (before) and zero times in the proposed system (after). This means that the operating system prevents the LMK operation from running to retrieve free memory whenever the memory capacity is less than the threshold of free physical memory.

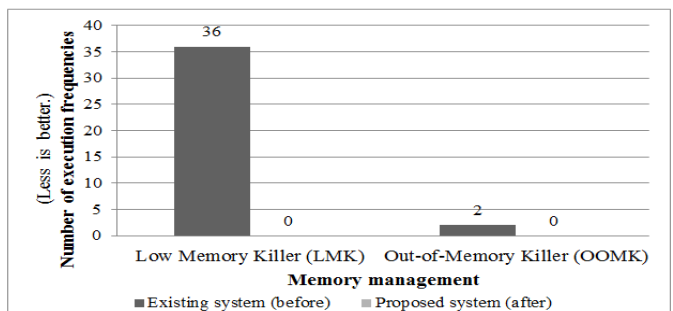

**Fig. 6. Number of execution frequencies of LMK/OOMK**

Figure 6 also compares the number of executions of OOMK between the existing system (before) and the proposed system (after). If the operating system cannot obtain the available memory with LMK, OOMK runs to avoid an out-of-memory situation. If the operating system cannot finally also gain the available memory with the OOMK operation, the out-of-memory problem results in system panic and system reboot. From the experiments, the number of OOMK operations is 2 times in the existing system (before) and zero times in the proposed system (after).

*C. Effect of Memory Defragmentation*

If the execution and termination of applications occur frequently, memory fragmentation increases. This section shows a comparison of the improved result of the memory fragmentation between the existing system (before) and the proposed system (after). When applications run initially, the operating system allocates the necessary memory space via a memory allocator [34]. If free memory blocks are smaller than the memory space actually needed, the operating system cannot allocate a memory space for new applications. As a result, the operating system must forcibly execute the activity to terminate the running processes. This is called an external memory fragmentation. Examples of such memory fragmentation in mobile devices are as follows:

1) The external memory fragmentation results from a large kernel data structure. At this time, the allocation using the *vmalloc* kernel function is an exception.
2) The external memory fragmentation results from the contiguous memory allocation for the peripheral devices.
3) The external memory fragmentation results in too many small memory blocks. The cost of memory management to control many small blocks, increases.

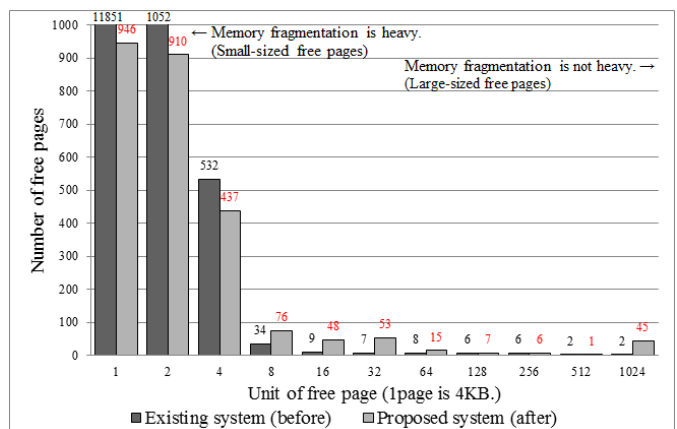

**Fig. 7. Comparison result of memory fragmentation between the existing system and the proposed system**

Figure 7 shows a comparison of the experimental result of external memory fragmentation between the existing system (before) and the proposed system (after). The value of the x-axis refers to the size unit of the memory block. The size of one memory page is 4 KB in the experimental condition. The size of the page depends on the CPU architecture. The value of the y-axis refers to the number of free pages. If the number of the y-axis increases, the external memory fragment [28], [26] is poor. The number on the left side of the x-axis indicates that memory fragmentation is heavy, while the number on the right side of the x-axis shows that memory fragmentation is not heavy.

The proposed system exhibited less memory fragmentation than the existing system [26]. The number of *1 free page* that refers to small-sized free page decreased from 11,851 (before) to 946 (after). The number of *1024 free page* that refer to large-sized free page increased from 2 (before) to 45 (after).

Figure 8 shows the percentage of memory fragmentation that occurs in the trusted part (VNODE0) and the untrusted part (VNODE1), corresponding to the two virtual nodes. The experimental result shows that the trusted part is isolated from the increased memory fragmentation problem of the untrusted part due to virtual memory partitioning. In addition, the trusted applications of VNODE0 can always run instantly because the complete memory isolation protects the trusted applications from the memory pressure of the untrusted applications.

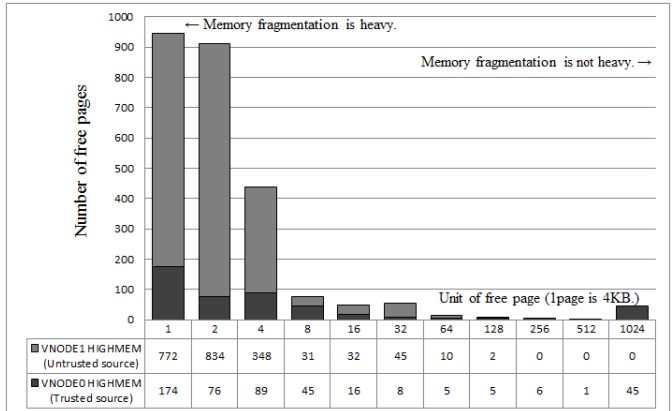

**Fig. 8. Percentage of memory fragmentation in the two virtual nodes**

*D. Execution Time of the Core Applications*

The experimental result demonstrates the execution time of two built-in applications to verify the effect of the real application by the improvement of LMK/OOMK and the minimization of memory fragmentation. The two built-in applications are as follows:
1) Phone application: The phone application is the most frequently used application among the built-in applications.
2) SMS application: The short message service is the second most frequently used application on a mobile phone. SMS is a text messaging service application of a phone.

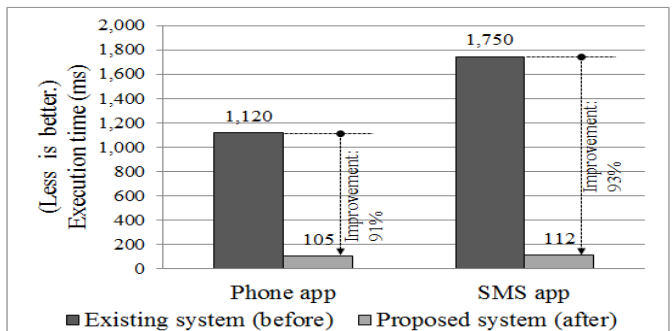

**Fig. 9. Execution time of the built-in applications (VNODE0)**

Figure 9 shows the execution result of the applications: the Phone application was improved by about 91% from 1.120 seconds (before) to 0.105 seconds (after) and the execution time of the SMS application was improved by about 93% from 1.750 seconds (before) to 0.112 seconds (after). The execution time of the SMS application in the existing system was 1.750 seconds because of the processing time needed for the database to read the SMS data. Therefore, the execution time of the SMS application depends on the performance of the database.

Figure 10 shows the breakdown for the core built-in applications. The overall degradation for each application consists of memory fragmentation, LMK, and OOMK. The y-axis shows the fraction of the total performance degradation that each factor causes. Since factors that cause page reclamation on a real device overlap in complex and integrated ways, it is not possible to obtain a precise separation. These results are an approximation that is intended to direct attention to the true thrashing [9] in the system.

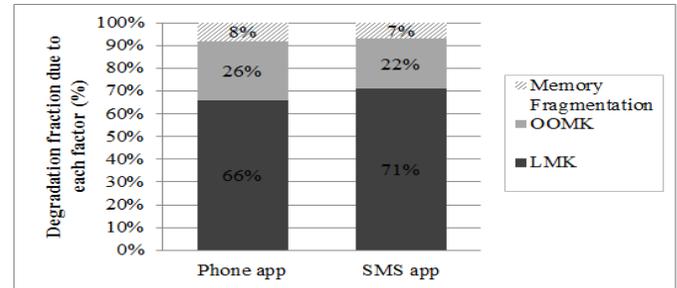

**Fig. 10. Contribution of each factor to the performance degradation**

While the proposed technique assures the instant user responsiveness of the time-critical trusted applications by using a novel virtual memory partitioning scheme, it has two limitations: memory utilization and performance of untrusted applications. The unknown applications forming the untrusted source cannot use the free memory space of the trusted part (VNODE0), even though the unknown applications in the untrusted part (VNODE1) need more free memory space. In this case, the trusted part does not need the many extra memory space because of the statically fixed built-in applications such as Phone, SMS, and Contacts in the trusted part (VNODE0). Therefore, the proposed scheme does not incur the fine tuning problem in which the initial limits in a real environment need to be determined.

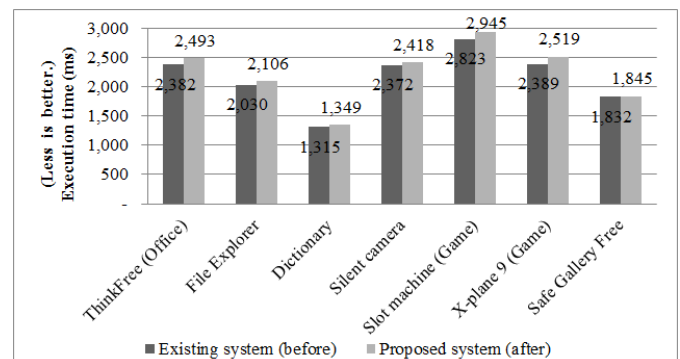

**Fig. 11. Execution time of the untrusted applications (VNODE1)**

The performance of applications from untrusted sources (VNODE1) could be degraded in the proposed approach, since the approach allocates memory within VNODE1. Figure 11 compares the performance of the downloaded applications in two configurations: without VNODE and with VNODE. It shows that the reduction in the performance of the untrusted applications is less than 3%. The approach is practical and useful because it can improve the performance of the trusted application by over 90% with a small, 3%, performance degradation of untrusted application.

## V. RELATED WORK

The memory resource controller [21] controls the memory quantity of tasks by limiting the memory usage of the task group managing the processes. The task group is a set of tasks to provide process aggregations in the operating system for resource tracking purposes. The approach helps the operating system that controls the memory usage of a task group with logical memory grouping based on the task's characteristics (e.g. web applications, media applications, office applications, and game applications). The user-space container can easily aggregate many threads with the memory resource controller. The user-space container is the user-space control package used to provide a user-space container object that provides full resource isolation and resource control for applications. However, this approach cannot settle memory fragmentation because logical memory groupings are used to intersect the area of one physical memory. Moreover, their system cannot isolate the operation of the page reclamation of untrusted applications including malware [2] and memory hog [34].

The Mondriaan Memory Protection (MMP) [35] proposed by E. Witchel improves the stability and maintainability of the kernel by supporting memory isolation using a permission key-based protection domain for the effective memory guard among the multi-domains that share the linear address space. However, this method depends on hardware because CPU needs to check the permission for effective address management based on a *protection lookaside buffer* (PLB). Therefore, MMP is not suitable for existing mobile devices due to the cost of the entire modification from the hardware architecture to the operating system.

S. Nomura [1] proposed a new selection policy of process termination by a relaunch, because the existing selection based on the policy does not always meet users' requirements. The proposed policy terminates processes with a short launching time. This approach can avoid the large time consumption of applications with long launching times. However, this technique cannot cover the frequent operation situation of LMK when the available memory is short. This approach focuses on fair user responsiveness by considering the loading time of the application. That is, it does not handle the fast execution time of the trusted application.

P. Barham [36] proposed a hypervisor as the complete resource virtualization. In addition, this approach supports the functionality for resource isolation and high performance and enhances the stability of the kernel with memory protection and *para-virtualization* [37]. However, this technique focuses on the virtualization of resources to share the high-performance system such as a server without mobile devices. The performance of the application in the native operating system is better than that of the guest operating system due to the virtual machine manager. Above all, the cost of the virtual machine manager is very high because the virtual machine manager needs to guarantee the performance of the applications in the guest operating systems. In other words, this approach does not support a lightweight memory isolation solution at the operating system level.

## VI. CONCLUSION

The conventional memory management features frequently induce thrashing, page fault, and page replacement to secure free memory. The proposed virtual memory node mechanism inhibits the performance degradation of applications caused by trashing, frequent page faults, and page replacements [7], [9]. It minimizes page reclamation of time-critical built-in applications from the trusted sources and limits the memory access range of unknown applications from the untrusted sources. By using the dynamic memory-controlling interface, different memory layouts can be configured at the boot time according to device types, such as mobile phones, tablets, laptops, and camcorders. In addition, the proposed approach supports complete virtual memory isolation based on a discontiguous memory access model to separately run applications from the trusted sources [2], [27] and the untrusted sources. It drastically reduces the number of LMK/OOMK operations by reducing the number of page faults and page replacements [7]. Consequently, the proposed approach overcomes the low performance of the trusted applications induced by LMK/OOMK operations during memory shortage.

## BIOGRAPHIES

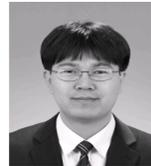

**Geunsik Lim** (S'13) received his B.S. degree in Computer Science and Engineering from Ajou University, in Korea in 2003. He is currently an M.S. student in the College of Information and Communication Engineering, Sungkyunkwan University, and a senior software engineer for Samsung Electronics in Korea. His current research interests include system optimization, embedded operating systems, mobile platforms, and multicores.

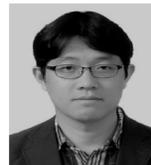

**Changwoo Min** received his B.S. and M.S. degrees in Computer Science from Soongsil University, Korea in 1996 and 1998, respectively. He is currently a Ph.D. candidate in the College of Information and Communication Engineering, Sungkyunkwan University, and a software engineer for Samsung Electronics in Korea. His current research interests include virtualization, storage systems, and mobile platforms.

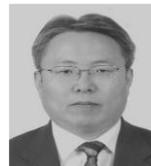

**Young Ik Eom** received his B.S., M.S., and Ph.D. degrees in the Department of Computer Science and Statistics of Seoul National University, Korea in 1983, 1985, and 1991, respectively. From 1986 to 1993, he was an Associate Professor at Dankook University in Korea. He was also a visiting scholar in the Department of Information and Computer Science at the University of California, Irvine, from Sep. 2000 to Aug. 2001. Since 1993, he has been a professor at Sungkyunkwan University in Korea. His research interests include virtualization, operating systems, cloud systems, and system securities.